\documentstyle{l-aa}
\def\o2{[OII]~$\lambda 3727$}
\def\h{~$h^{-1}$ Mpc~}

\def\hmagp{$+5\log h$~}

\begin{document}
\thesaurus{ 11(11.04.1; 11.12.2; 12.03.3; 12.12.1) }
\title{ The ESO Slice Project (ESP) galaxy redshift survey:
\thanks{Based on observations collected at the European Southern
Observatory, La Silla, Chile.} }
\subtitle{ I. Description and First Results }
%
%-------------------------------------------------------------------------------
%
\author{
G.Vettolani\inst{1}
\and
E.Zucca\inst{2,1}
\and
G.Zamorani\inst{2,1}
\and
A.Cappi\inst{2}
\and
R.Merighi\inst{2}
\and
M.Mignoli\inst{2}
\and
G.M.Stirpe\inst{2}
\and
H.MacGillivray\inst{3}
\and
C.Collins\inst{4}
\and
C.Balkowski\inst{5}
\and
V.Cayatte\inst{5}
\and
S.Maurogordato\inst{5}
\and
D.Proust\inst{5}
\and
G.Chincarini\inst{6,7}
\and
L.Guzzo\inst{6}
\and
D.Maccagni\inst{8}
\and
R.Scaramella\inst{9}
\and
A.Blanchard\inst{10}
\and
M.Ramella\inst{11}
}
%
%-------------------------------------------------------------------------------
%
\institute{ 
Istituto di Radioastronomia del CNR, 
via Gobetti 101, 40129 Bologna, Italy
\and
Osservatorio Astronomico di Bologna, 
via Zamboni 33, 40126 Bologna, Italy
\and
Royal Observatory Edinburgh, 
Blackford Hill, Edinburgh EH9 3HJ, United Kingdom
\and
School of EEEP, Liverpool John--Moores University, 
Byrom Street, Liverpool L3 3AF, United Kingdom
\and
Observatoire de Paris, DAEC, Unit\'e associ\'ee au CNRS, D0173 et \`a
l'Universit\'e Paris 7, 5 Place J.Janssen, 92195 Meudon, France
\and
Osservatorio Astronomico di Brera, 
via Bianchi 46, 22055 Merate (LC), Italy
\and
Universit\`a degli Studi di Milano, 
via Celoria 16, 20133 Milano, Italy
\and
Istituto di Fisica Cosmica e Tecnologie Relative, 
via Bassini 15, 20133 Milano, Italy
\and
Osservatorio Astronomico di Roma, 
via Osservatorio 2, 00040 Monteporzio Catone (RM), Italy
\and
Universit\'e L. Pasteur, Observatoire Astronomique, 
11 rue de l'Universit\'e, 67000 Strasbourg, France
\and
Osservatorio Astronomico di Trieste, 
via Tiepolo 11, 34131 Trieste, Italy
}
%
%-------------------------------------------------------------------------------
%
\offprints{Elena Zucca (zucca@astbo1.bo.cnr.it)}
\date{Received 00 - 00 - 0000; accepted 00 - 00 - 0000}
\maketitle
\markboth {G.Vettolani et al.: 
The ESP galaxy redshift survey: I. Survey Description}{}
%
%-------------------------------------------------------------------------------
%
\begin{abstract}
The ESO Slice Project (ESP) is a galaxy redshift survey we have recently
completed as an ESO Key--Project. The ESP covers 23.3 square degrees in a 
region close to the South Galactic Pole.  The survey is nearly complete 
(85\%) to the limiting  magnitude $b_J=19.4$ and 
consists of 3342 galaxies with  reliable redshift determination.

In this paper, the first in a series that will present the results of the 
ESP survey, we describe the main characteristics
of the survey and briefly discuss the properties of the galaxy sample. 
From a preliminary spectral analysis of a large sub--sample of 2550
galaxies we find that the fraction of actively star--forming galaxies
increases from a few percent for the brightest galaxies up to about 40\% for
the galaxies fainter than $M= -16.5$ \hmagp. 

The most outstanding feature in the ESP redshift distribution is a very
significant peak at $z \simeq 0.1$. The detection of similar peaks, at the
same distance, in other surveys in the same region of the sky,
suggests the presence of a large
bidimensional structure perpendicular to the line of sight. The minimum size
of this structure would be of the order of $ 100 \times 50$ \h,
comparable with the size of the Great Wall.

\end{abstract}
%%%%%%%%%%%%%%%%%%%%%%%%%%%%%%%%%%%%%%%%%%%%%%%%%%%%%%%%%%%%%%%%%%%%%%%%%%%%%%%%
\section{Introduction}

In the course of the last decade redshift surveys have 
provided a major advance in our knowledge
of the large scale distribution of galaxies and its statistical properties
(see for example the review of Giovanelli and Haynes 1991).

Bright, wide angle surveys (e.g. CfA2, Geller and
Huchra 1989; SSRS2, da Costa et al. 
1994; Perseus--Pisces, Giovanelli and Haynes 1988)
cover a large fraction of the sky and provide a clear 
picture of the nearby universe, up to about 10,000 km/s
or $100$ \h (where $h = H_0 / 100$). 
These surveys reveal large structures in the distribution of galaxies:
voids of sizes up to $50$ \h (de Lapparent et al. 1986), and 
large ($>100$ \h) bidimensional
sheets as for instance the Great Wall within CfA2 (Geller and Huchra 1989), 
the Southern Wall within SSRS2 (da Costa et al 1994), or the Perseus--Pisces
filament (Giovanelli and Haynes 1988).

Strategies alternative to those adopted for the above--mentioned surveys
have been used to study the large--scale 
structure in depth without paying the price of an
excessively large increase of the observing time.
These strategies include: a) sparse 
sampling (see Kaiser 1986), as in the Stromlo--APM redshift survey
(Loveday et al. 1992); b) chessboard surveys consisting of separated fields 
covering a large solid angle; c) pencil beam surveys, as in the 
Broadhurst et al. (1990; BEKS) survey. These deeper surveys
have confirmed the texture detected in shallower surveys 
up to a depth of 40000 -- 50000 km/s.

These strategies, while allowing a faster completion of the surveys,
sometimes do not allow an unambiguous interpretation of the data.
For example, BEKS report evidence of periodic structures
in their first pencil beam redshift survey. However, the real nature of 
this periodicity is not clear. Moreover, the periodicity is not confirmed 
in other directions of the sky by the same authors
(Koo 1993) and is not detected in other similarly deep
redshift surveys even in directions close to the original BEKS
pencil beam (Bellanger and de Lapparent 1995).

The present survey (hereafter ESP: ESO Slice Project) was designed 
to provide an unbiased spectroscopic sample of galaxies brighter than
$b_J = 19.4$, with a high level of completeness, over
a region of the sky significantly extended in the right ascension direction.
The geometry of the survey, a slice,  
is the most efficient for mapping three-dimensional 
structures like those observed in shallower surveys (de Lapparent et al. 1988),
provided its thickness is not smaller than the correlation length of the galaxy
distribution. Our sample can be used to derive the
basic statistical
properties of the galaxy distribution, averaged over a volume large enough to 
smooth out ``local'' inhomogeneities. With a magnitude limit $b_J \leq 19.4$
we include $L^*$ galaxies up to $z \simeq 0.16$ and obtain a redshift 
distribution peaking at $z \simeq 0.1$. Because of the target selection
criteria, the ESP is unbiased with  respect to a number of statistical
descriptors, such as, for example, the luminosity function (see Zucca
et al. 1997, paper II).

The redshift distribution of the ESP galaxies is similar to that of the
galaxies in the Las Campanas Redshift Survey (LCRS, Shectman et al. 1996),
which covers over 700 square degrees in six strips, each $1.5^o \times 80^o$,
and consists of 26418 redshifts of galaxies.
The comparison between the two surveys, however, is not 
straightforward. In fact, ESP galaxies are selected in the photographic blue
band, while LCRS galaxies are selected on red CCD frames. Furthermore,
contrary to what was done for the LCRS, we did not apply any ``a priori''
selection on the surface brightness of the target galaxies. 

%
%------------------------------------------------------------------------------
% FIGURE 1. 
\begin{figure*}
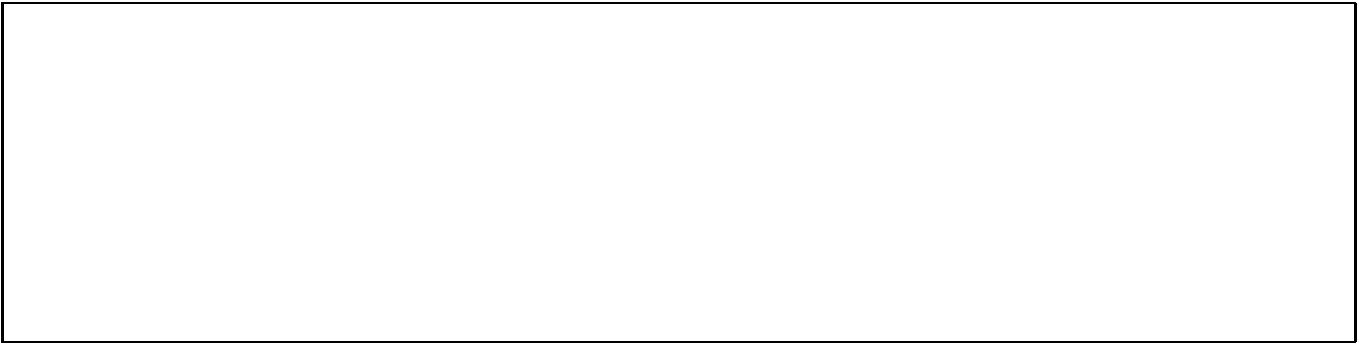

\picplace{4.5cm}
\caption[]{  Survey geometry. }
\end{figure*}
%------------------------------------------------------------------------------
%

In Section 2 we briefly describe the survey region, the galaxy catalogue, 
the observations and the data reduction; in Section 3 we discuss the main
spectral properties of the galaxies in our sample;
in Section 4 we describe qualitatively the large--scale structures detected
in our survey. We summarize the main properties of the ESP redshift
survey in Section 5.

%%%%%%%%%%%%%%%%%%%%%%%%%%%%%%%%%%%%%%%%%%%%%%%%%%%%%%%%%%%%%%%%%%%%%%%%%%%%%%%%
\section{General description of the survey}

A detailed description of the survey will be presented in the paper containing
the whole data catalogue (Vettolani et al. 1997, paper III, in preparation).
Here we briefly summarize the main characteristics of the ESP project.

%%%%%%%%%%%%%%%%%%%%%%%%%%%%%%%%%%%%%%%%%%%%%%%%%%%%%%%%%%%%%%%%%%%%%%%%%%%%%%%%
\subsection{The photometric sample}

The main area of the ESP survey is a strip $22^o$ long in right ascension
and $1^o$ thick in declination (hereafter strip A). In order to make full
use of the allotted nights, we were able to survey also an area
of $5^o \times 1^o$ (hereafter strip B), five degrees west of strip A.
Both strips are located in the region of the South Galactic Pole. 
The position was chosen in order to minimize the galactic absorption effects 
($-60^o \la b^{II} \la -75^o$). The right ascension limits are from 
$22^{h} 30^m$ to $ 22^{h} 52^m$ for strip B and from $23^{h} 23^m$
to $01^{h} 20^m$ for strip A, at the mean declination 
$\delta = -40^o 15'$ (1950).

The target galaxies, with a limiting magnitude $b_J = 19.4$, 
were extracted from the Edinburgh--Durham Southern Galaxy 
Catalogue (EDSGC, Heydon--Dumbleton et al. 1988, 1989), 
which has been obtained from COSMOS scans of SERC J survey plates. 
The EDSGC has a $95\%$ completeness 
at $b_J \leq 20.0$ and an estimated stellar contamination $\leq 10\%$ 
(Heydon--Dumbleton et al. 1989). 

Preliminary analysis of CCD data, obtained with the 0.9m Dutch/ESO telescope
for about 80 galaxies in the magnitude range
$16.5 \leq b_J \leq 19.4$ in the region of our survey, 
shows a linear relation between $b_J$(EDSGC) and $m_B$(CCD), 
with a dispersion ($\sigma_M$) of about 0.2 
magnitudes around the fit (Garilli et al. in preparation). 
Since the CCD pointings cover the entire right ascension range of our 
survey, this $\sigma_M$ includes both statistical errors within single
plates and possible plate--to--plate zero point variations.

We do not have enough information to assess the reliability
of magnitudes for the 71 ESP galaxies with $m < 16.5$.
It is known that in this bright range various problems may affect
the measures, as saturation, ghosts and spikes on plates, the presence of 
substructure in bright galaxies, or contamination from overlapping bright 
stars (e.g. Loveday 1996).
With this {\em caveat} in mind, we simply note that presently we do not have 
evidence for large errors on the magnitudes of bright galaxies in our 
catalogue, on the basis of a few cases we could check.

%-------------------------------------------------------------------------------
% TABLE 1.
\begin{table}
\caption[]{ Sample statistics }
\begin{flushleft}
\begin{tabular}{lllllll}
\hline\noalign{\smallskip}
Sample & Catalogue & Observed & Galaxies  & Stars & Failed \\
\noalign{\smallskip}
       & Objects & Objects    & with z  &   &Spectra \\
\noalign{\smallskip}
\hline\noalign{\smallskip}
%%%%%%%%%%%%%%%%%%%%%%%%%%%%%%%%%%%%%%%%%%%%%%%%%%%%%%%%%%%%%%%%%%%%%%%%%%%%%%%%
strip A   & 3391 & 3265 & 2749 & 368 & 147 \\
strip B   & 1096 & ~779 & ~593 & 125 & ~61 \\
total     & 4487 & 4044 & 3342 & 493 & 208 \\
%%%%%%%%%%%%%%%%%%%%%%%%%%%%%%%%%%%%%%%%%%%%%%%%%%%%%%%%%%%%%%%%%%%%%%%%%%%%%%%%
\noalign{\smallskip}
\hline
\end{tabular}
\end{flushleft}
\end{table}
%-------------------------------------------------------------------------------
%
%------------------------------------------------------------------------------
% FIGURE 2. 
\begin{figure*}
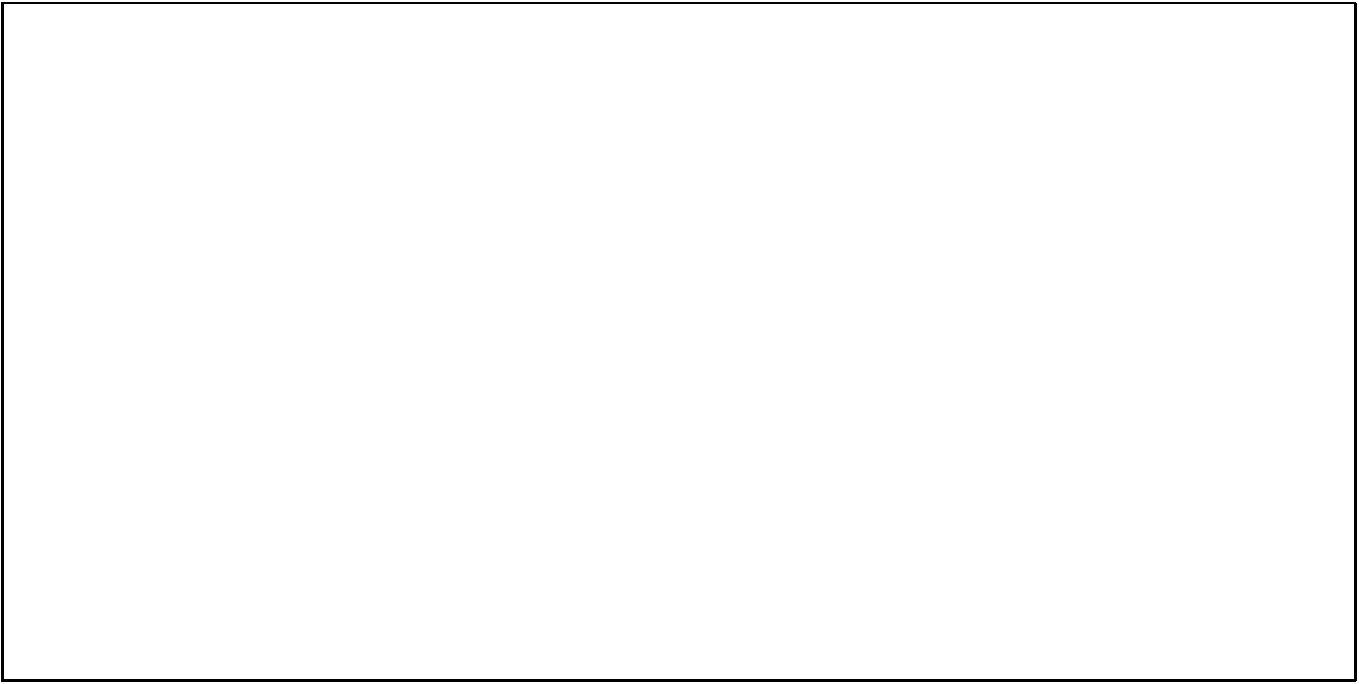

\picplace{9.0cm}
\caption[]{Redshift completeness for each OPTOPUS field. Field numbers 
increase with right ascension. 
a) Northern row.
Field numbers from 1 to 9 correspond to region B; 
field numbers from 21 to 65 correspond to region A.   
b) Southern row.
Field numbers from 101 to 109 correspond to region B; 
field numbers from 121 to 164 correspond to region A.
}
\end{figure*}
%------------------------------------------------------------------------------

%%%%%%%%%%%%%%%%%%%%%%%%%%%%%%%%%%%%%%%%%%%%%%%%%%%%%%%%%%%%%%%%%%%%%%%%%%%%%%%%
\subsection{Observations and data reduction}

The observations have been obtained during six observing runs (in the period
1991--1993) with the multifiber spectrograph OPTOPUS (Lund 1986, Avila et al. 
1989) at the Cassegrain focus of the ESO 3.6m telescope at La Silla. 
The geometry of the survey (see Figure 1) is
a regular grid consisting of two rows of 
adjacent circular fields. Each field has a diameter of 32 arcmin,
corresponding to the field of view of the OPTOPUS spectrograph.
The two rows of fields are separated
by 15 arcmin and slightly overlap each other.
The total solid angle is $\sim 7.1 \times 10^{-3}$ steradians 
(i.e. 23.3 square degrees). 

The instrument OPTOPUS has 50 fibers, each with a diameter projected 
on the sky of $\sim 2.5$ arcsec. The fibers
are manually plugged into holes drilled in aluminum plates at the galaxy 
positions. For each field  we reserved at least 5 fibers for the
measurement of the spectrum of the blank sky. 
On average, there are 35  galaxies brighter than $b_J = 19.4$ 
per OPTOPUS field and $80 \%$ of the fields have fewer than 45 galaxies. 
When the number of galaxies in a field exceeded the number of 
available fibers, we selected at random the galaxies 
to be observed from the total
galaxy list. Whenever possible, we re-observed overdense fields 
in order to reach a higher completeness.

Another set of observations has been obtained 
in October 1994 with the multifiber 
MEFOS spectrograph (Felenbok et al. 1997). Thanks to these
observations we have been able to observe some of the
galaxy spectra that had no redshift determination 
after the OPTOPUS runs. In particular, we have 
used these observations to reach a completeness as uniform as possible
over all the OPTOPUS fields in strip A.

The spectra cover the wavelength range from 3730\AA~to 6050\AA, with 
an average pixel size of 4.5\AA. 
We measured the redshifts of the galaxies by cross--correlating 
the sky-subtracted spectra with a set of 8 
template stars. The template stars have been observed with 
the same instrumental set-up used to obtain the galaxy spectra. 
We also measured redshifts from the emission lines, whenever present.
The median internal velocity error is of the order of $\sim 60$ km/s.
From a comparison of our 8 templates with three SAO radial velocity
standard stars we estimate that the zero--point error should be smaller
than $\sim 10$ km/s. We will give a full description of our 
observations and our reduction procedure in paper III.

%%%%%%%%%%%%%%%%%%%%%%%%%%%%%%%%%%%%%%%%%%%%%%%%%%%%%%%%%%%%%%%%%%%%%%%%%%%%%%%%
\subsection{Sample statistics}

We observed a total of 4044 objects, corresponding to $\sim 90\%$ of the 
parent  photometric sample of 4487 objects.
Out of the 4044 observed objects, 493 turned out to be stars and 
208 have a too low signal--to--noise ratio to provide a reliable redshift
(failed spectra). In the end, our final sample consists of a total of 
3342 galaxies with reliable redshifts ($+$ one quasar). About half of the 
galaxies in our sample have detectable emission lines (see Section 3).
In Table 1 we report a summary of the basic numbers relative to  
strip A and strip B.

The mean completeness of our galaxy  sample is $\sim 85\%$. 
We derive this value by assuming that all the failed spectra 
correspond to galaxies and that the percentage of stars among the 
443 objects that were not observed is the same as 
in the spectroscopic sample (i.e. $\sim 12\%$).
The completeness level is significantly different in 
strip A and in  strip B, being $91\%$ for strip A and $64\%$ for strip B. 
The different completeness level between the two strips is due to 
our choice of repeating observations only for fields in strip A.
Figure 2 shows the completeness for each OPTOPUS field: panel a) refers to the 
northern row (field numbers $<100$), panel b) to the southern row (field 
numbers $>100$).

A detailed understanding of the completeness and the selection effects is 
extremely important in any analysis of our data. We therefore performed
a number of tests to assess the statistical properties of the galaxies 
for which we did not measure the spectrum or we did not get a measurable 
spectrum. 

First of all, in the fields where the number of objects is higher than
the number of available fibers, the objects we observed are a  
random subset of the total catalog with respect to both
magnitude and surface brightness. The set of observed galaxies departs
from randomness only with respect to the selection of 
the position of the target objects, due to the instrumental constraint
that two fibers can not be put closer than about 25 arcsec; obviously
this means that when dealing with pairs of galaxies
at small projected separation, we could observe
only one galaxy (except when the other galaxy was selected
for a second exposure). 

As far as failed spectra are concerned,
we find that the fraction of failed spectra
increases at fainter magnitude. However, the fractional increase
of failed spectra at fainter magnitudes is not strong,
reaching  $\sim 7$\%  at the limit $b_J = 19.4$.
This effect is taken into account in our statistical analysis
of the survey (see for example paper II).

%%%%%%%%%%%%%%%%%%%%%%%%%%%%%%%%%%%%%%%%%%%%%%%%%%%%%%%%%%%%%%%%%%%%%%%%%%%%%%%%
\section{The spectral properties of the sample}

The ESP data provides a large data-base of
galaxy spectra which, when fully analyzed, will constitute a reference for 
the next generation of larger and deeper redshift surveys that will probe
galaxy evolution ($z \geq 0.3$). These data are well suited for the study
of the intrinsic properties of galaxies in terms of their stellar populations,
and their relation with the environment and redshift (cosmic time). This
will be presented in a future paper.

Emission lines are present in a large fraction ($\sim 50\%$) 
of the galaxies in our sample. We mainly detect [OII] $\lambda 3727$, 
$H\beta$, [OIII] $\lambda 4959$ and $\lambda 5007$.
Our preliminary analysis focused on 
the \o2~doublet, since this line is the most useful star formation tracer 
(Kennicutt 1992) in the wavelength range covered by our spectra.
Galaxies showing the \o2~line correspond, at increasing equivalent width, to
three main categories: spiral galaxies, where the line originates
mostly from HII regions in the disks, galaxies undergoing
a significant burst of star formation, or AGNs.

Great care is required when dealing with statistical analysis
of emission lines properties of galaxies, because the detectability of lines
in a spectrum strongly depends on the signal--to--noise ratio $S/N$ of the 
adjacent continuum.
The $S/N$ ratio in the blue part of our spectra ranges from
$\sim$ 2 to $\sim$ 10, with an average value of the order of 4. With such
a value the \o2~ line can be detected only if its equivalent width is
larger than  about 5 \AA. For the spectra of poorer quality, however, 
the minimum
detectable \o2~ equivalent width is of the order of 20 \AA~(see, for example,
Figure 7 in Vettolani et al. 1994). For this reason, in order
to properly analyze the 
distribution of the equivalent widths below 20 \AA~and to study the possible
correlations of the equivalent width with other intrinsic properties, one 
should also take into account the information carried by the upper limits.
In this paper, in order to avoid dealing with upper limits, 
we present a preliminary analysis of the galaxies with an \o2~ equivalent
width greater than 20 \AA.

Up to now the measurement of the equivalent width of the \o2~line,
or of an upper limit for the galaxies in which the line is not
detected, has been done for a subsample of 2550 galaxies, corresponding
to 76\% of the total sample.
From this subsample we find that about 13\% of the galaxies show an
\o2~equivalent width greater than 20 \AA~and can therefore be classified
as actively star--forming galaxies. This fraction, however, is a strong
function of absolute magnitude, ranging from a few
percent for the brightest galaxies up to about 40\% for
the galaxies fainter than $M= -16.5$ \hmagp (see Figure 3). This result is 
qualitatively similar to what has been found in the AUTOFIB Redshift Survey
(see Figure 8 in Ellis et al. 1996). Obviously, this correlation has
to be taken into account in the determination of the evolution of the star 
formation rate as estimated by comparing bright and local samples of
galaxies with fainter and more distant ones. 

%------------------------------------------------------------------------------
% FIGURE 3.
\begin{figure}
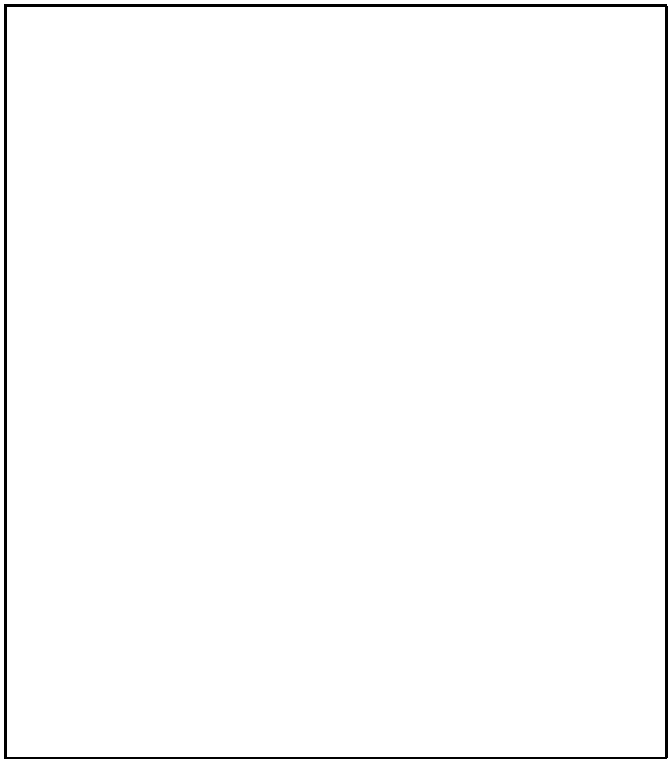

\picplace{10cm}
\caption[]{ Percentage of galaxies showing [OII] $\lambda 3727$ with
an emitted equivalent width greater than 20 \AA~as a function of absolute 
magnitude.}
\label{fig:histoemi}
\end{figure}
%-------------------------------------------------------------------------
%
%-----------------------------------------------------------------------------
%  FIGURE 4. 
\begin{figure*}
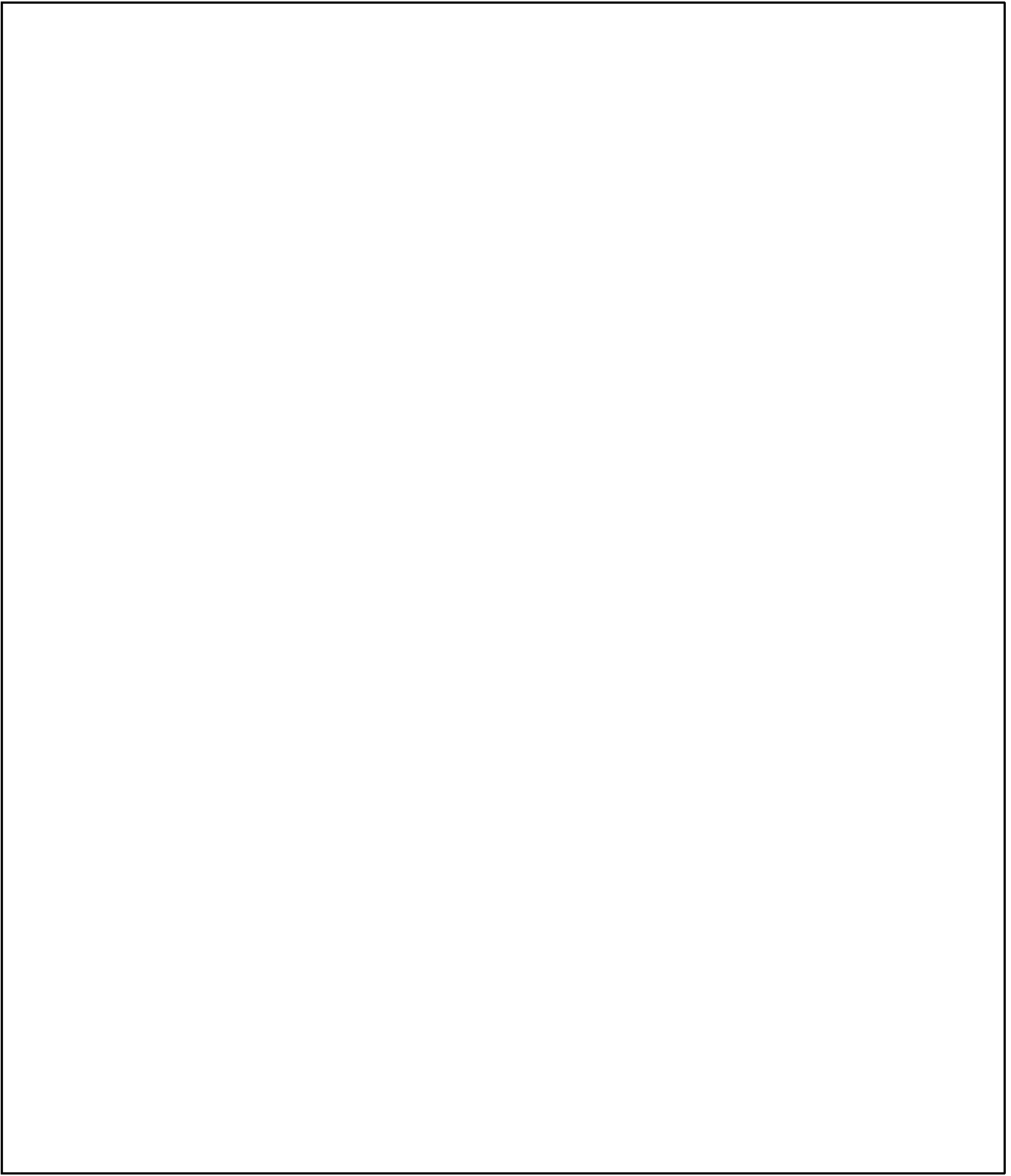

\picplace{21.0cm}
\caption[] {~a) Galaxy distribution in comoving distance ($q_0 = 0.5$); error
bars represent $1\sigma$ Poissonian uncertainties. 
The solid line shows the expectation resulting from a uniform distribution of 
the galaxies in the sample. Vertical lines correspond to the positions of the 
BEKS peaks (see text). b) Wedge diagram of ESP galaxies projected in right 
ascension; the step of the grid is $30^m$. 
c) The same as panel b) for galaxies in the LCRS strip centered at 
$\delta=-39^o.$}
\end{figure*}
%
%------------------------------------------------------------------------------

%%%%%%%%%%%%%%%%%%%%%%%%%%%%%%%%%%%%%%%%%%%%%%%%%%%%%%%%%%%%%%%%%%%%%%%%%%%%%%%%
\section{The large--scale structure}

One of the main goals of our survey is to study the properties 
of the large--scale structure in the local Universe.
Our survey is significantly deeper than local surveys (e.g. CfA2 \& SSRS2) 
since $M^*$ galaxies are sampled out to $z \simeq 0.16$, corresponding
to a comoving distance $d_{com} \sim 440$ \h ($q_0 = 0.5$), which is 
the effective depth of the sample.
This depth, together with a wide extension in the right ascension 
direction ($22^o$ corresponding to $\sim 145$ \h at the
effective depth), could be enough to ensure
the characteristic of a "fair sample" to ESP. Clearly, the
discovery of structures larger than those observed in the existing 
shallower surveys may prevent even ESP from being a fair sample.
 
In particular, ESP is deep enough to detect at least the
first peaks of the BEKS survey and it is sufficiently wide
and thick ($\sim 6.5$ \h at $z=0.16$) to clarify the nature of the 
peaks that could correspond either to isolated groups/clusters or
to the intersection of the beam with a larger connected structure like
the Great Wall.

Detailed statistical results, including an analysis of the groups and
clusters in the survey, will be presented in forthcoming papers;
here we describe the main characteristics of the volume sampled
by the ESP.
 
Figure 4 shows the histogram of the distribution in comoving distance
of the 3342 galaxies with measured redshift (panel a) and
the corresponding wedge diagram (panel b). 
For comparison we show in panel c) the wedge diagram corresponding
to the ESP right ascension range from the nearest strip of the
LCRS centered at $\delta = -39 ^o$ (Shectman et al. 1996).

The most outstanding feature
in Figure 4a is the peak at $\simeq 290$ \h. Although about 50 redshifts in 
this peak are due to galaxies within 1 Abell radius from the centers of two 
Abell clusters (ACO 2840 and ACO 2860) both located near the eastern edge of 
strip A, this peak remains highly significant even if these galaxies are 
removed from the sample. In fact, it is clear from Figure 4b that 
the peak in the redshift distribution
corresponds to an ensemble of large--scale structures nearly perpendicular 
to the line of sight. If this ensemble can be considered as a single,
connected structure, it would have a linear dimension of about 120 \h,
before entering into the gap between regions A and B.
At least part of this structure is visible in the LCRS data (panel c)
where it seems to extend towards west and possibly connecting to other 
structures.

Other foreground/background peaks in the redshift histogram
correspond to structures less evident to the eye in the
wedge diagram. 

Close examination of the wedge diagram shows the existence of many more
dense regions and structures, some of which are elongated along the line 
of sight such as the one in region B stretching from about 130 \h to 200 \h. 

The LCRS data in panel c) allow to follow
some of these structures, even if the different sampling, color selection
and depth of the samples do not allow a quantitative comparison.

As well as all the other redshift surveys (see for instance panel c),
our data suggest the presence of large underdense regions. Figure
4a clearly shows one such region at $d_{com} \simeq 225$ \h and a second one
corresponding to a nearby ($d_{com} \leq 140$ \h) underdensity.

%
%------------------------------------------------------------------------------
% FIGURE 5.
\begin{figure}
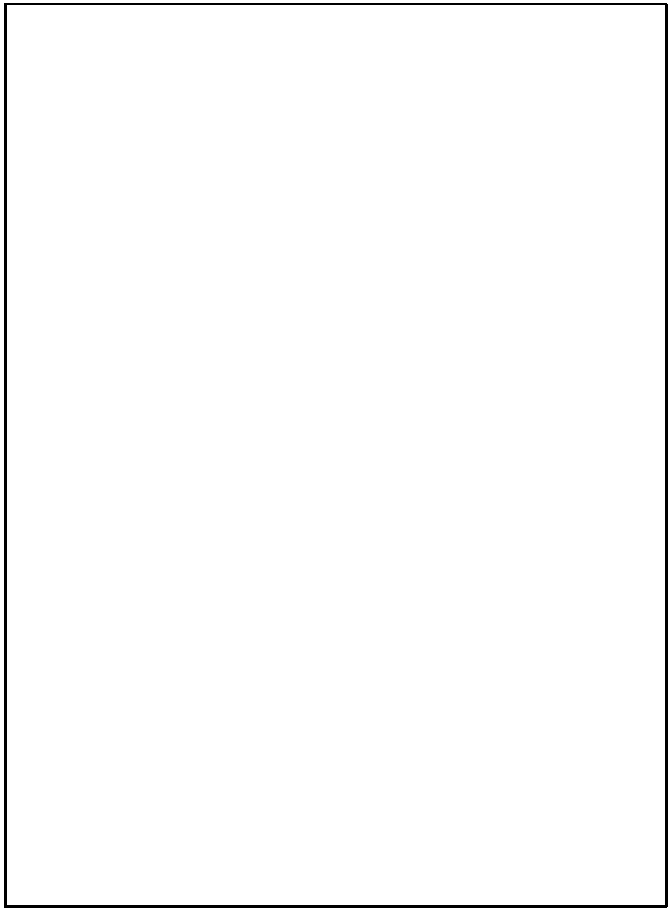

\picplace{12.0cm}
\caption[]{ Smoothed galaxy distribution, obtained by taking into account the 
survey selection function and the fields incompleteness (see text). }
\end{figure}
%------------------------------------------------------------------------------

On the basis of our data and the comparison with the LCRS data,
we conclude that this nearby underdense region is statistically
significative (see Figure 4a); however,
given the small solid angle covered by ESP and LCRS
in this region, it is impossible to assess its extension in the transverse 
direction. This underdensity has interesting
consequences for the interpretation of the bright galaxy counts as we
discuss in paper II. 

The vertical lines in Figure 4a show the location of the regularly 
spaced density enhancements found in the BEKS pencil beam survey, 
located at $\alpha = 00^h 54^m$, $\delta = -27.9^o$, i.e.
$\sim 12^{\circ}$ north of the eastern part of the ESP.
The two main peaks in our redshift distribution (at comoving distances
of $\sim 170$ and $\sim 290$ \h) are in reasonably good agreement 
with the BEKS peaks. The same two peaks are clearly visible in the
LCRS data extracted from their slice immediately north ($\delta \sim - 39^o$)
of our strip. Moreover, the peak at $\sim 290$ \h appears to be present
in the deep pencil beam of Bellanger and de Lapparent (1995),
centered at $\alpha = 00^h 19^m$ and $\delta = -30.25^o$, i.e.
$\sim 7^o$ away from the BEKS direction.
It is also visible in the shallower OPTOPUS probes obtained in this 
region of the sky by Ettori et al. (1995, see their Figure 2).

The coincidence in depth of the two nearest peaks of the redshift distribution
within ESP with similar peaks in all other surveys in the same 
region suggests the presence of large 
extended structures (walls), approximately orthogonal to the
line of sight. Under this hypothesis, the structure at $z \simeq 0.1 $ would
have minimum linear dimensions of the order of $ 100 \times 50$ \h,
comparable with the size of the Great Wall (Geller and Huchra 1989).

The existence of these two peaks does not imply, however, a
periodicity with a preferred scale as claimed by Broadhurst et al. (1990), 
which can be tested only with much deeper samples.

Figure 5 shows the density of the {\em planar projection} of the data. 
Isodensity contour levels are spaced as $2^j$, with $j=0,1,...$:
they represent the ratio $n_d/n_r$ of the ESP projected number density 
$n_d$, Gaussianly smoothed over $5$ \h, to the projected
number density $n_r$ of a similarly smoothed random field, 
derived from an average of 100 $3D$ random samples.

Each galaxy was given a weight inversely proportional to
the selection function (as derived from the ESP luminosity function, see 
paper II), and to the incompleteness of its field.
This procedure implicitly assumes that the galaxies we did not observe 
follow the same distribution in depth as the observed galaxies;
this assumption is justified by our random selection of targets in overdense 
fields.
The weighting with the selection function allows to detect 
structures even at large distances, which would be otherwise ``washed out'';
however, as any magnitude--limited sample, such structures are inferred from 
the few high luminosity galaxies 
that can still be detected, and are therefore affected by a large 
uncertainty.

The random samples were built using the ESP
average luminosity function and K--corrections described in paper II, and
within a three--dimensional space which
faithfully takes into account the sample geometry, i.e.
not only the unobserved region, but also
the presence of a few ``holes" in the 
original catalogue, due to the presence of bright stars.

The final result is a representation of the large--scale galaxy distribution
which is more ``objective'' than the cone--diagram.

In Figure 5 the dense regions corresponding to the peaks in Figure 4a
are clearly identifiable, in particular the two structures crossing the entire
region from one side to the other at comoving distances of 
of $\sim 170$ and $\sim 290$ \h. Note also that the most prominent structure 
at large comoving distance in the small strip B appears to be elongated along 
the line of sight.

%%%%%%%%%%%%%%%%%%%%%%%%%%%%%%%%%%%%%%%%%%%%%%%%%%%%%%%%%%%%%%%%%%%%%%%%%%%%%%%%
\section{Summary}

We have presented the main characteristics of the ESO Slice Project,
a redshift survey limited to $b_J = 19.4$ in a region near the South Galactic 
Pole. 

This survey, consisting of 3342 galaxies with reliable redshift determination,
provides a large data-base of galaxy spectra which will
be useful to study the spectral properties of galaxies and will constitute
a reference for the next generation of larger and deeper redshift surveys.
Preliminary analysis of a large sub--sample of 2550 galaxies shows
a strong anti--correlation between the fraction of galaxies with 
an \o2 equivalent width greater than 20 \AA~and the absolute magnitude. 
This fraction of actively star--forming galaxies
increases from a few percent for the brightest galaxies up to about 40\% for
the galaxies fainter than $M= -16.5$ \hmagp. 

Finally, this survey allows the study of the large--scale structure
at intermediate depth ($z \sim 0.16$).
The most outstanding feature in the redshift distribution is a very
significant peak at $\simeq 290$ \h.  Similar peaks, at the same
distance, are detected in other surveys in the same region of the sky.
We conclude that there is evidence for the presence of a large
bidimensional structure perpendicular to the line of sight. The minimum size
of this structure would be of the order of $ 100 \times 50$ \h,
comparable with the size of the Great Wall.

%%%%%%%%%%%%%%%%%%%%%%%%%%%%%%%%%%%%%%%%%%%%%%%%%%%%%%%%%%%%%%%%%%%%%%%%%%%%%%%%

\begin{acknowledgements}
This work has been partially supported through NATO Grant CRG 920150,  
EEC Contract ERB--CHRX--CT92--0033, CNR Contract 95.01099.CT02 and by 
Institut National des Sciences de l'Univers and Cosmology GDR.
\\
It is a pleasure to thank the support we had from the ESO staff both in 
La Silla and in Garching. In particular, we are grateful to Gerardo Avila for 
his advice and his help in solving every instrumental problem we have been
facing during this project. 
\\
Finally, we thank the referee for his constructive comments.
\end{acknowledgements}

%%%%%%%%%%%%%%%%%%%%%%%%%%%%%%%%%%%%%%%%%%%%%%%%%%%%%%%%%%%%%%%%%%%%%%%%%%%%%%%%

%%%%%%%%%%%%%%%%%%%%%%%%%%%%%%%%%%%%%%%%%%%%%%%%%%%%%%%%%%%%%%%%%%%%%%%%%%%%%%%%

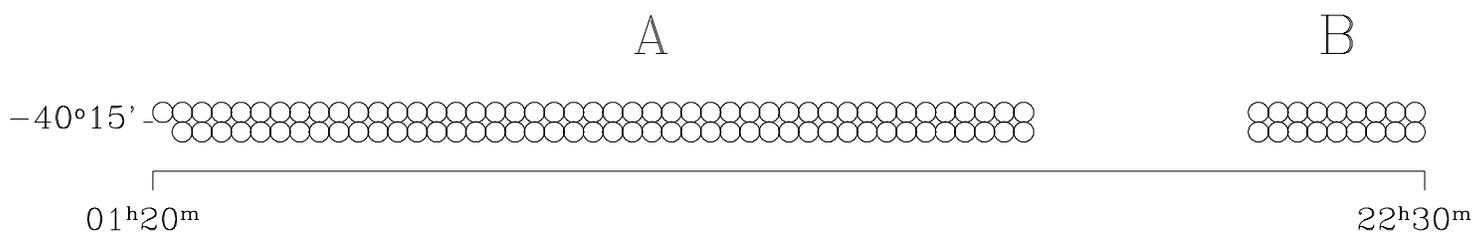
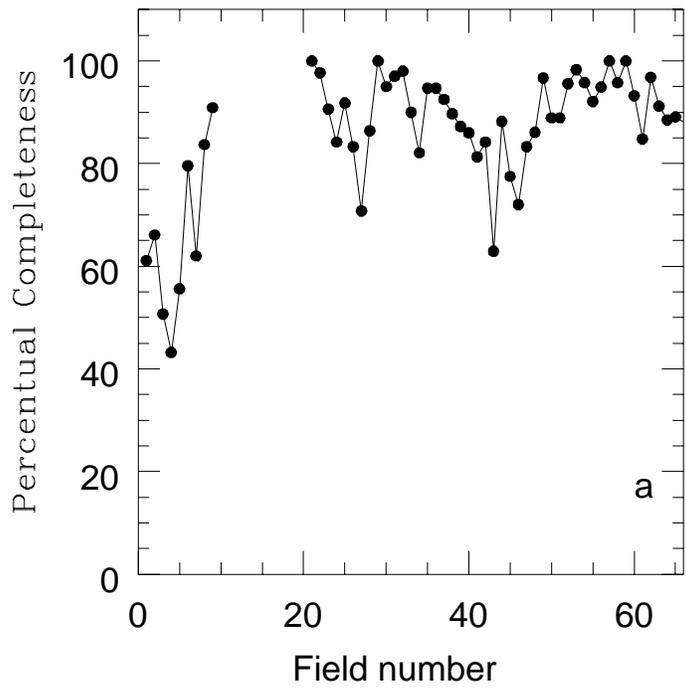
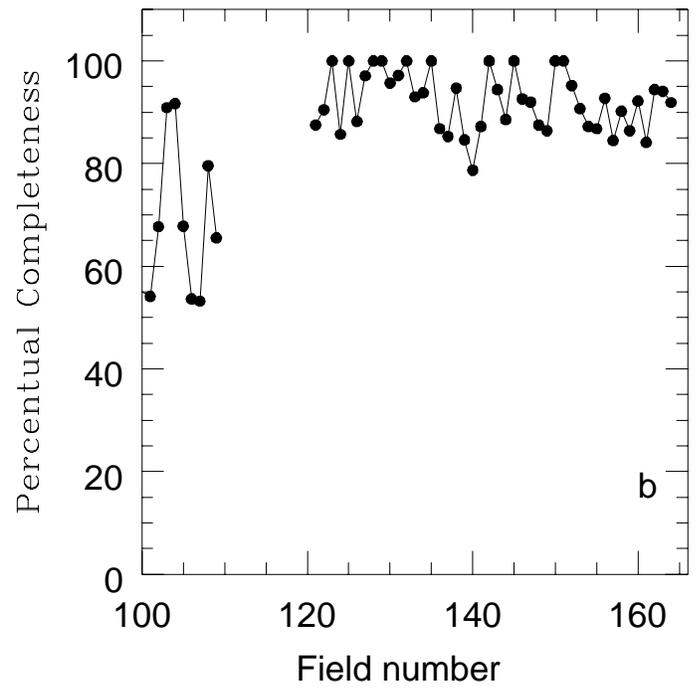
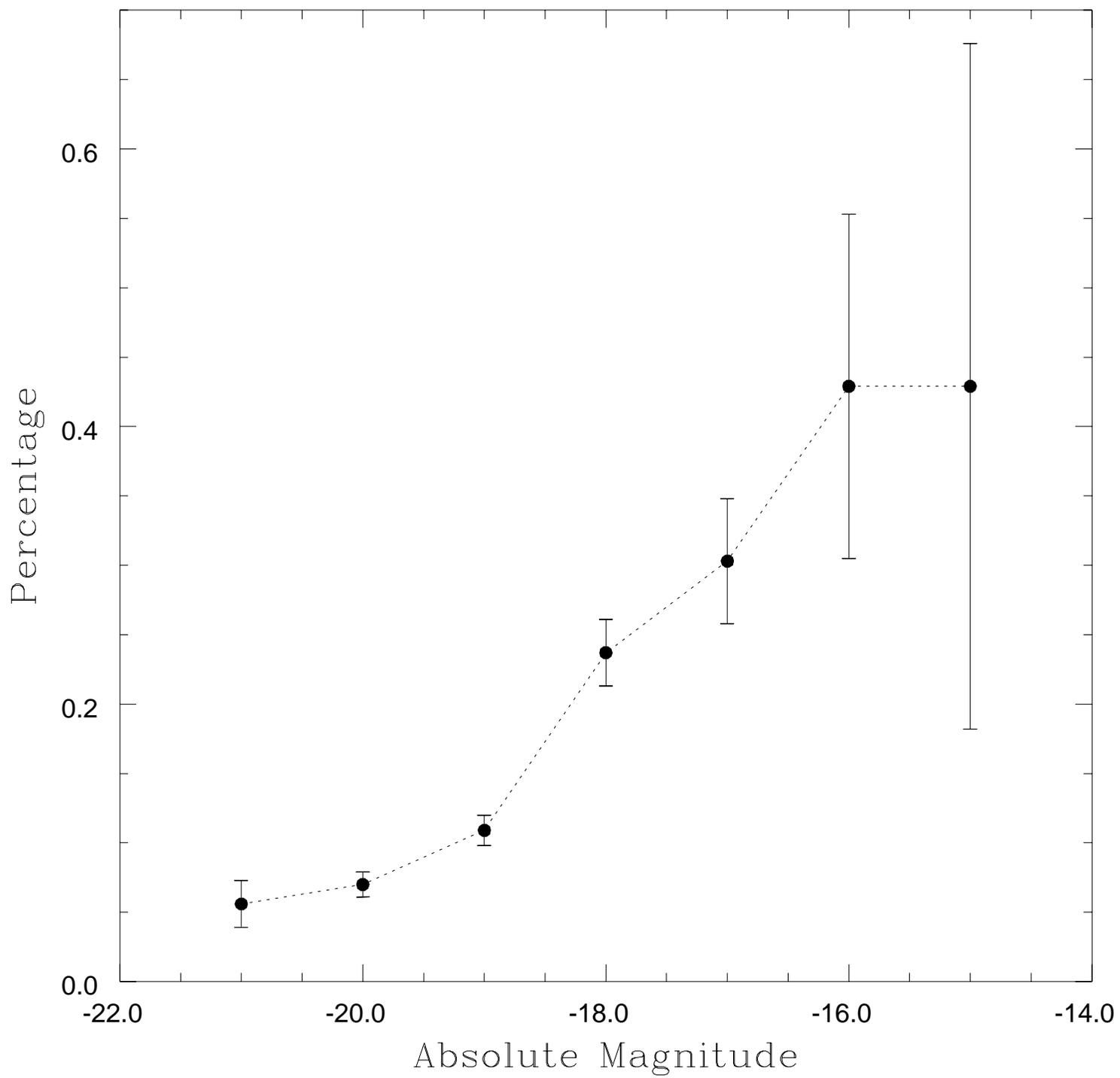
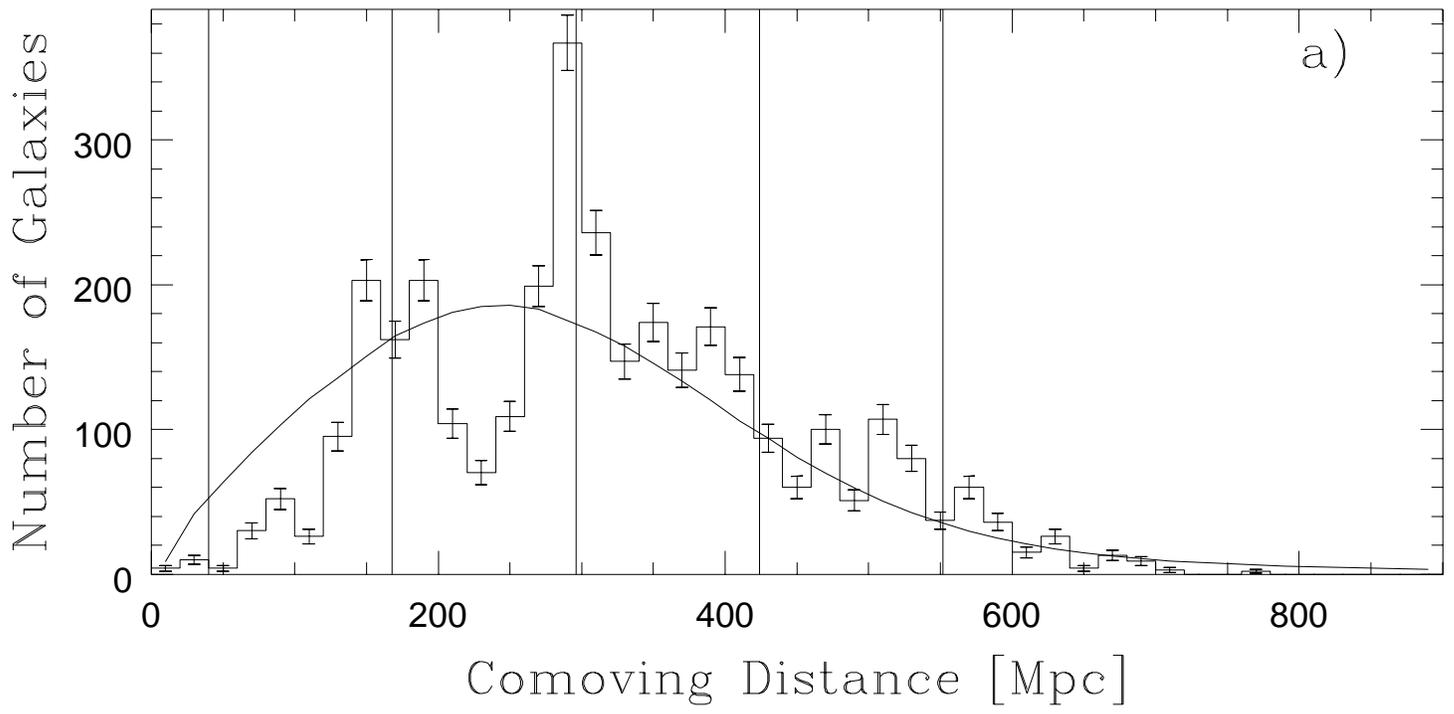
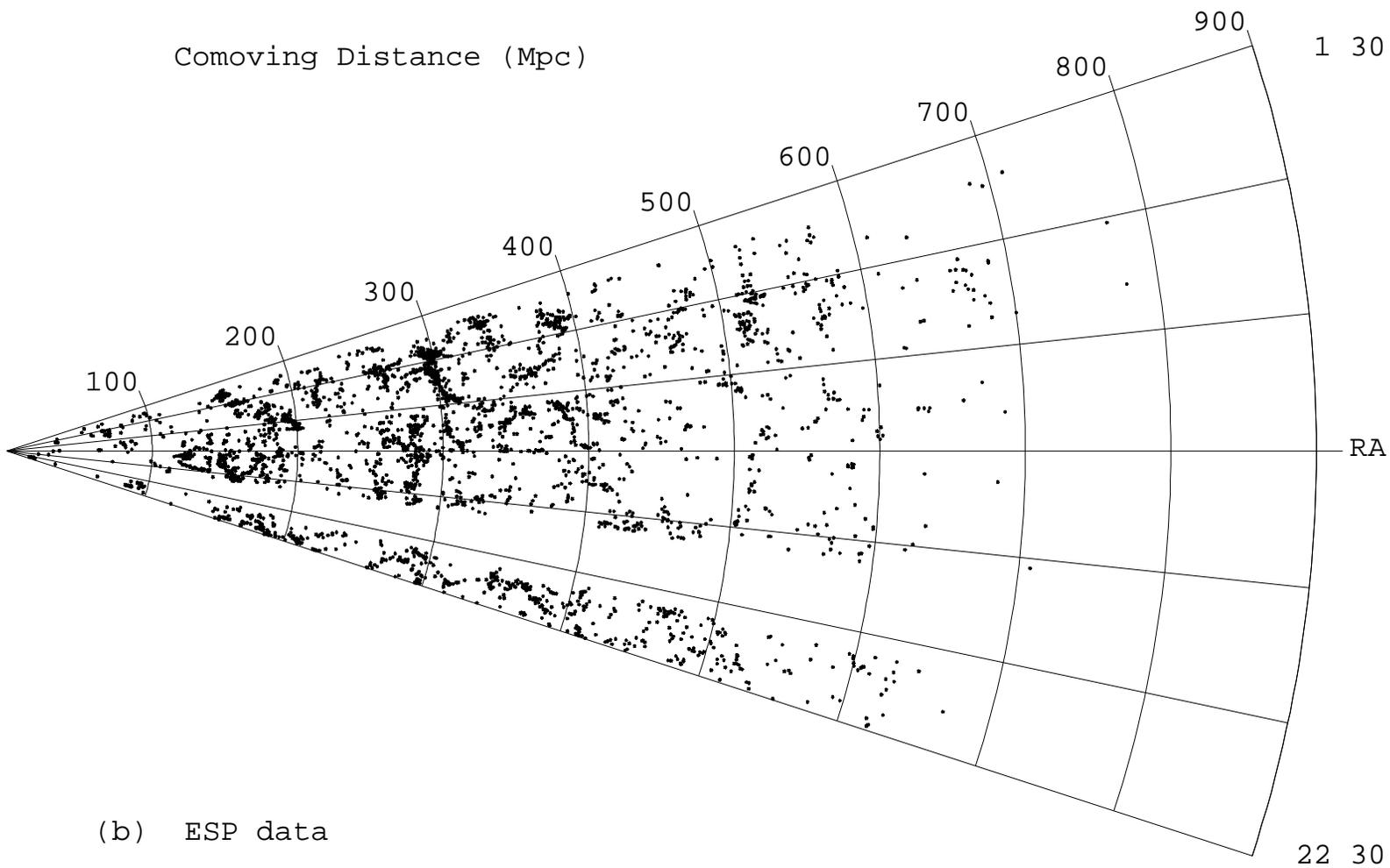
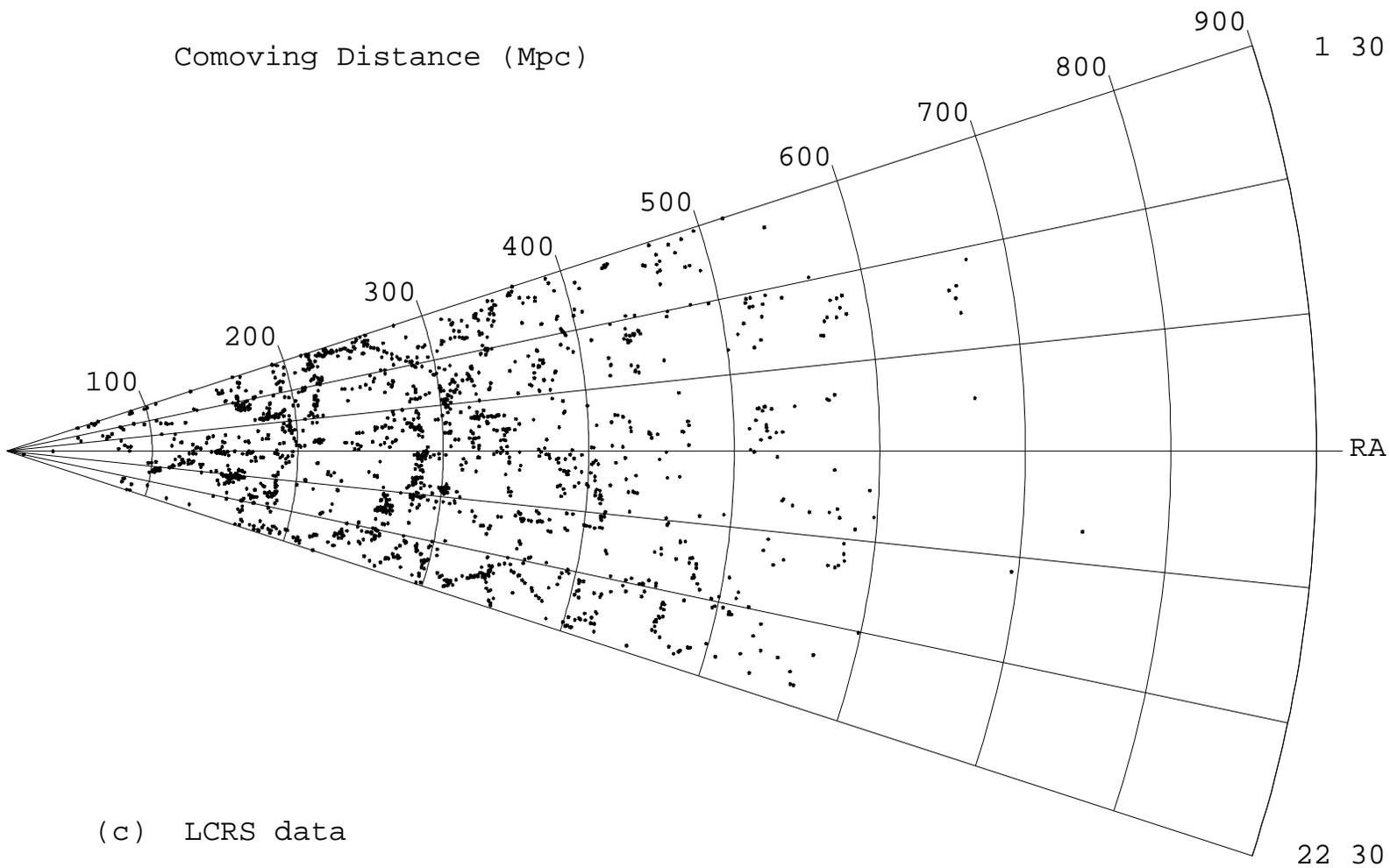
\end{document}